\documentclass{IOS-Book-Article}
\usepackage{amsmath}
\usepackage{mathptmx}
\usepackage{graphicx}

\usepackage{soul}\setuldepth{article}
\def\hb{\hbox to 11.5 cm{}}

\begin{document}

\pagestyle{empty}
\thispagestyle{empty}
\def\thepage{}
\begin{frontmatter}              

\title{Including the Cost of Irreducible Uncertainty in the Policy Compression Framework}

\markboth{}{March 2026\hb}

\author[A]{\fnms{\'Alvaro} \snm{Garrido-P\'erez}\orcid{0009-0003-5481-8166}
\thanks{Corresponding Author: \'Alvaro Garrido-P\'erez, al(dot)garridoperez(at)ugent.be}},
\author[A]{\fnms{Pieter} \snm{Simoens}\orcid{
0000-0002-9569-9373}}
and
\author[A]{\fnms{Amrapali} \snm{Pednekar}\orcid{0009-0005-6194-3955}}
and
\author[B]{\fnms{Yara} \snm{Khaluf}\orcid{orcid.org/0000-0002-5590-9321}}
\runningauthor{\'A. Garrido-P\'erez et al.}
\address[A]{IDLab, Department of Information Technology, Ghent University - imec, Ghent, Belgium}
\address[B]{Network Institute, Vrije Universiteit Amsterdam, Amsterdam, Netherlands}

\begin{abstract}
AI decision-support systems can benefit from anticipating biases in human decision-making. Many such biases may arise from human cognitive limitations. The policy compression framework models decision-making as a trade-off between reward maximization and the cognitive cost of encoding state-dependent action policies, formalized as the mutual information between states and actions (policy complexity). We argue that this account is incomplete because it treats conditional entropy--the irreducible uncertainty about which action should be selected given a state--as costless, even though empirical evidence suggests that it modulates reaction times. We therefore extend the framework by defining cognitive cost as the sum of policy complexity and a weighted conditional-entropy term, governed by
a new parameter, $\eta$. The resulting optimal policy retains the standard exponential form but becomes sharper as $\eta$ increases, allowing policy precision to vary more independently of reward sensitivity. This modification implies that the standard policy compression framework may underestimate the cognitive cost of action selection, and it has the potential to better account for biases in human decision-making. At the same time, it introduces additional complexity for fitting the model to human data, which future work will need to address.

\end{abstract}

\end{frontmatter}
\markboth{March 2026\hb}{March 2026\hb}

\section{Introduction}

Humans must continually select appropriate actions in response to changing states of the world. At a simplified level, this process can be understood as a mapping from states ($s$) to actions ($a$), or more generally to a probability distribution over actions. In reinforcement learning (RL), this mapping is formalized as a policy, denoted by ($\pi(a \mid s)$). In standard RL, the goal is to learn a policy that maximizes expected reward, where the expected value of taking action ($a$) in state ($s$) is given by the \textit{value function} ($Q(s,a)$).

However, representing and choosing a policy is not free. Humans have limited cognitive and memory resources, and storing and retrieving information about a specific policy may come at the expense of processing other vital information. As a result, decision-making is shaped by a reward-complexity trade-off: more complex policies can produce higher rewards, but they also require greater memory resources. This trade-off may lead to systematic biases or incorrect decisions, which may be particularly harming in high-stakes settings such as driving, piloting, or medical decision-making. Understanding how humans manage this trade-off is therefore relevant not only for cognitive science, but also for AI decision-support systems that aim to anticipate when human decisions are likely to become unreliable.

The resource-rational framework proposes that humans solve the aforementioned trade-off optimally \cite{lieder2020resource}, meaning that we choose the best possible action subject to our cognitive constraints. From this perspective, understanding human decision-making requires answering two questions:  how are rewards or preferences learned, and what is the cognitive cost of implementing a given policy? The first question is commonly addressed using RL, in which an agent learns action values through trial and error. The second has been addressed by the Policy Compression framework \cite{lai2021policy}, a computational modelling framework that has recently been used to explain a diverse range of decision-making phenomena, including perseveration \cite{gershman2020origin}, cognitive deficits in schizophrenia \cite{gershman2021reward}, and mouse navigation \cite{amir2020value}, among others \cite{liu2025time, gershman2025policy}.

The Policy Compression (PC) framework defines the cognitive cost of a policy in terms of how much information about states is used to select actions. Specifically, the cost is defined as the mutual information between states and actions, which is referred to as \textit{policy complexity} and can be expressed as follows:

\begin{equation}
    I^{\pi}(S;A)
=\sum_s P(s)\sum_a \pi(a\mid s)\log\frac{\pi(a\mid s)}{P(a)}=\sum_s P(s) D_{\text{KL}}[\pi(a\mid s)||P(a)]
\label{eq:policy_complexity}
\end{equation}
Where $P(a) = \sum_{s}P(s)\pi(a|s)$ is the marginal probability distribution over actions, that is, the average likelihood across all states of selecting a given action, and $D_{\text{KL}}[\cdot \mid \mid  \cdot]$ is the Kullback-Leibler divergence.

\begin{figure}[h!]
    \centering
    \includegraphics[width=1\textwidth]{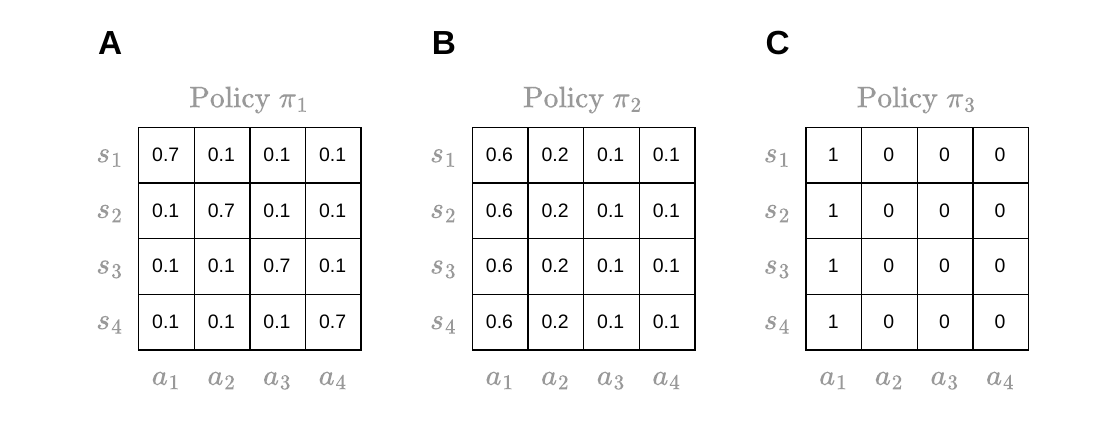}
    \caption{Examples of three policies: $\pi_1$ (\textbf{A}), $\pi_2$ (\textbf{B}), and $\pi_3$ (\textbf{C}). For any non-deterministic state distribution (i.e., any distribution other than one in which a single state has $P(s_i)=1$ and all others have probability zero), policy $\pi_1$ has non-zero policy complexity, whereas policies $\pi_2$ and $\pi_3$ always have zero policy complexity.}
    \label{fig:policies_examples}
\end{figure}

Intuitively, when policy complexity is large, the probability of selecting a given action depends strongly on the state in which we are. This incurs a memory cost because we must store information about these contingencies. For example, in Fig. \ref{fig:policies_examples}\textbf{A} we see an example of a policy ($\pi_1$) where there are four possible states and four possible actions, and each state has an associated (and different) action that is most probable. Thus, ideally, we would need to store all the information presented in the table. In contrast, in Fig. \ref{fig:policies_examples}\textbf{B}, we see an example of a policy ($\pi_2$), with a complexity equal to zero because all states are associated with the same distribution over actions. Therefore, we only need to store one vector that can be used across all states, resulting in a lower cognitive cost, since we do not need to store or retrieve any information regarding states.

The definition of cognitive cost (Eq. \eqref{eq:policy_complexity}) was directly drawn from rate-distortion theory, a branch of information theory concerned with how systems can compress information under limited capacity while tolerating some loss of fidelity \cite{cover1991rate}. Under this analogy, decision-making is like a noisy communication channel: the state of the world is the input, the selected action is the output, and the cognitive limitations are the \textit{channel capacity} which determines how much state information can be transmitted to select an action.

In addition to defining policy complexity as a cognitive cost, the PC framework assumes that policy selection is constrained by limited cognitive resources. A capacity-limited agent can therefore be described as selecting the policy that maximizes expected reward ($V^{\pi}$) while keeping policy complexity equal to a fixed capacity bound C, resulting in the following optimization problem:

\begin{equation}
\max_{\pi}\;\; \beta V^{\pi}-I^{\pi}(S;A)
\label{eq:pol_comp_constrained}
\end{equation}
subject to
\[
I^{\pi}(S;A) = C .
\]

where two other necessary constraints (namely, that action probabilities must be non-negative and sum to 1) are left implicit and the expected reward of a given policy $\pi$, is given by $V^{\pi} = \sum_{s}P(s)\sum_{a}\pi(a\mid s)Q(s,a)$.
 
The solution to the problem (i.e., the optimal policy) takes the following form \cite{lai2021policy}:

\begin{equation}
    \pi^{*}(a\mid s) \propto \exp[\beta \, Q(s,a)+ \log P^{*}(a)]\label{eq:optimal_policy_compression}
\end{equation}

where the inverse temperature parameter $\beta$ determines how much the policy depends on maximizing rewards in contrast to staying close to the marginal action distribution or \textit{baseline} policy $P^{*}(a)=\sum_s P(s)\pi^{*}(a\mid s)$.

\section{Why Policy Complexity may be an Incomplete Definition of Cognitive Cost}

Despite the aforementioned successes of the PC framework, we argue that its definition of cognitive cost--namely, policy complexity--may be incomplete. To better illustrate our point, we can express policy complexity more intuitively in terms of two entropies:

\begin{equation}
    I^{\pi}(S;A)
=H(A)-H(A \mid S)\label{eq:policy_complexity_entropies}
\end{equation}

Where:

\begin{equation}
    H(A)=-\sum_a P(a) \log P(a)
    \label{eq:marginal_action_entropy}
\end{equation}

\begin{equation}
    H(A\mid S) = -\sum_s P(s)\sum_a \pi(a \mid s)\log \pi(a \mid s)=\sum_s P(s) H[\pi(a\mid s)]
    \label{eq:conditional_entropy}
\end{equation}

$H(A)$ is the \textit{marginal action entropy}, which determines the overall uncertainty about actions.  If $H(A)$ is small, then most of the probability mass of the policy will be concentrated on a few actions; when $H(A)$ is large, it will be more spread out. Thus, $H(A)$ can also be interpreted as a measure of the effective size of the action repertoire.

$H(A\mid S)$ is the \textit{conditional entropy}, which expresses the uncertainty about actions after observing the state, averaged over the state distribution. If $H(A\mid S)$  is large, then, on average, the probability distribution over actions for each state will be more diffuse.

Policy complexity is therefore the difference between the total uncertainty about actions, captured by $H(A)$, and the \textit{irreducible} (or \textit{aleatoric}) uncertainty captured by
$H(A \mid S)$ \cite{arumugam2024bayesian}. The latter is considered irreducible because it cannot be eliminated even when the current state is known. A simple example is a fair coin toss: there is a 50\% chance of heads and a 50\% chance of tails. Even if we know the relevant state of the world--namely, that the coin is fair--this uncertainty (or entropy) cannot be reduced.

It is important to clarify that, by irreducible uncertainty, we mean the uncertainty that remains after the current state is known, given the agent’s present understanding of the world--or, equivalently, given its current \textit{world model}. In this sense, the term ``irreducible'' may be somewhat misleading, because such uncertainty may be reduced over time through learning. Indeed, one could argue that, with perfect knowledge of the world, no uncertainty would be truly irreducible. For example, in the coin-toss case, if we had complete knowledge of the initial conditions--such as the positions and trajectories of the surrounding air molecules, as well as the force and angle of the toss--we could, in principle, predict whether the coin would land heads or tails.

The problem of identifying cognitive costs solely with policy complexity becomes apparent when we compare policy $\pi_2$ (Fig. \ref{fig:policies_examples}\textbf{B}) with policy $\pi_3$ (Fig. \ref{fig:policies_examples}\textbf{C}). Both policies have zero policy complexity because, regardless of the state, the probability of choosing a given action is the same. Consequently, there is no state information available to guide action selection (i.e., there is no mutual information between states and actions). Under the PC framework, both policies are therefore treated as equally demanding. Yet this seems counterintuitive: under $\pi_{3}$, one only needs to store information about $a_1$, whereas under $\pi_{2}$, one must store information about all available actions, which would presumably increase memory demands, even if only slightly.

This result follows directly from the definition of cost as mutual information. By construction, it captures only the extent to which action selection depends on state information. Any residual uncertainty captured by $H(A \mid S)$--as in $\pi_2$--is assumed to impose no cognitive burden.
We hypothesize however, that such an assumption may be too strong in the context of human decision-making and that policy complexity alone may underestimate the cognitive demands of policies that remain highly stochastic despite being state-independent.

Our hypothesis is consistent with evidence showing that humans often rely on simplified categorical representations such as ``all or nothing'' and ``right or wrong'', over more nuanced representations that allow for gray areas and a wider range of categories. This tendency, known as \textit{dichotomous thinking}, has been associated with a lower level of cognitive ability \cite{mieda2021dichotomous}, suggesting that it may reduce cognitive demand, at the cost of simplifying the true structure in the environment.

\section{Introducing the Cost of Irreducible Uncertainty in the Policy Compression Framework}

To account for the potential cost of irreducible uncertainty, we propose a modification of the PC framework in which the cost of a policy, \(c^\pi\), depends not only on its policy complexity, \(I(S;A)\), but also on its conditional entropy, \(H(A\mid S)\):
\begin{equation}\label{eq:new_cost}
    c^\pi = I(S;A) + \eta\, H(A\mid S)
    = H(A) + (\eta - 1)H(A\mid S),
    \qquad \text{with } \eta \in [0,1).
\end{equation}

Here, \(\eta\) determines how costly it is to encode the average irreducible uncertainty (captured by the conditional entropy) relative to policy complexity. When \(\eta = 0\), the conditional entropy term carries no additional cost, and Eq.~\eqref{eq:new_cost} reduces to the original PC framework.

The second equality in Eq.~\eqref{eq:new_cost} highlights two possible ways of reducing policy cost. One is to increase \(H(A\mid S)\): if, on average, action selection becomes more uncertain within each state, then knowing the current state provides less information about which action will be chosen, and the mutual information \(I(S;A)\) decreases. The other is to decrease \(H(A)\): if the overall action distribution becomes less informative (i.e., the effective action repertoire is reduced), then the mutual information between states and actions also decreases. In the original PC framework, both strategies contribute equally because cost is defined as \(H(A)-H(A\mid S)\). In the present formulation, however, increasing \(H(A\mid S)\) is a less efficient way of reducing overall cost, because even though it lowers the amount of state information required for action selection, it also increases the irreducible uncertainty that must be encoded.

Equation~\eqref{eq:new_cost} can be expanded as follows:
\begin{align}\label{eq:new_cost_expanded}
    c^\pi
    &= \sum_s P(s)\, D_{\mathrm{KL}}\!\left[\pi(a\mid s)\,\|\,P(a)\right]
    + \eta \sum_s P(s)\, H[\pi(a\mid s)] \\
    &= \sum_s P(s)\,
    \underbrace{\left\{
    D_{\mathrm{KL}}\!\left[\pi(a\mid s)\,\|\,P(a)\right]
    + \eta\, H[\pi(a\mid s)]
    \right\}}_{c_s^\pi}.
\end{align}

Under this decomposition, \(c_s^\pi\) can be interpreted as the cost of the policy in state \(s\). This expression implies that, for a given state, two quantities increase the cost. The first is the degree to which the policy deviates from the baseline action distribution, as measured by \(D_{\mathrm{KL}}[\pi(a\mid s)\,\|\,P(a)]\). This is consistent with information-theoretic accounts of cognitive cost which define it as a function of how much the policy deviates from a prior or habitual policy \cite{zenon2019information,parr2023cognitive}. Although a habitual prior need not be identical to \(P(a)\), the marginal action distribution can still be interpreted as a baseline policy insofar as it influences action selection independently of the current state. Indeed, prior work has shown that relying on \(P(a)\) results in perseveration \cite{gershman2020origin}--that is, the tendency to repeat an action even after the state changes--a phenomenon often associated with strong priors.

The second term is the irreducible uncertainty of a policy given a state, \(H[\pi(a\mid s)]\), also known as \textit{action uncertainty}. This term may help capture discriminability effects, whereby choices between similarly valued options tend to be slower than choices between options that differ substantially in subjective value \cite{fontanesi2019reinforcement}. When options are similarly valued, \(\pi(a\mid s)\) should be more uniform and therefore have higher entropy, which in the present framework corresponds to a higher cost. This term may also help account for learning-related effects that are not captured by the standard PC framework alone \cite{lai2024human}.

Studies modelling human decision-making using the PC framework in contextual multi-armed bandits have shown that reaction times (RTs) are related to both \(\log \frac{\pi(a\mid s)}{P(a)}\), which reflects the deviation between  \(\pi(a\mid s)\) and \(P(a)\), and to the irreducible uncertainty of the policy given the state observed on that trial, \(H[\pi(a\mid s)]\) \cite{lai2024human,lai2025action}. However, in these studies, the latter term must typically be added \emph{ad hoc} when predicting RTs, because it is not part of the formal definition of cost (see also \cite{garrido2025cognitive}).

Although RTs are not a direct measure of cognitive cost, they are often used as a proxy (e.g., \cite{yao2025neural,lee2021trading}), based on the assumption that slower responses reflect more information processing. Thus, even if the processing rate is not strictly constant, a monotonic relationship between cognitive cost and RT should still be expected. This raises a natural question: if both the amount of deviation from $P(a)$ and the level of irreducible uncertainty in a policy affect RTs, why should only the former be considered part of the cognitive cost? Our proposal is to incorporate the latter directly into the cost function.

\section{Analysis of the Extended Optimal Policy}

Given the revised definition of cost, the optimal policy is obtained by maximizing expected value subject to the new cognitive constraint:

\begin{equation}
\max_{\pi}\;\; V^{\pi}
\label{eq:main_obj}
\end{equation}
subject to
\[
I(S;A) + \eta\, H(A\mid S) = C.
\]

As shown in the Appendix, the corresponding optimal policy takes the form:

\begin{equation}
\pi^*(a\mid s)=
\frac{
\exp\left\{\frac{1}{1-\eta}\left[\beta\,Q(s,a)+\log P^*(a)\right]\right\}
}{Z(s)}
\label{eq:policy_extended}
\end{equation}
where \(Z(s)\) is the partition function (see Appendix), and \(P^*(a)=\sum_s p(s)\pi^*(a\mid s)\).

The resulting optimal policy (Eq. \eqref{eq:policy_extended}) retains the same general exponential form as in the standard PC framework, but now depends on two parameters, $\beta$ and $\eta$. As in the original formulation (Eq. \eqref{eq:optimal_policy_compression}), $\beta$ plays a role analogous to an inverse-temperature parameter, controlling the trade-off between reward maximization and adherence to the marginal action distribution \(P^*(a)\). Higher values of $\beta$ place greater weight on reward, whereas lower values place greater weight on the tendency to select globally frequent actions.

The new parameter, $\eta$, determines the cost assigned to irreducible uncertainty and therefore controls the overall sharpness of the policy. Specifically, $\eta$ rescales the exponent by the factor $1/(1-\eta)$, which is always greater than or equal to 1 for $\eta \in [0,1)$. Consequently, any $\eta > 0$ results in a \textit{sharper} policy relative to the standard PC solution. In this sense, the original PC optimal policy can be understood as the highest-entropy limiting case of the present formulation, obtained when $\eta = 0$. Intuitively, once irreducible uncertainty is itself costly, the optimal policy becomes less diffuse and more certain or deterministic.

An important consequence of introducing this additional parameter is that policy precision and reward sensitivity become partially dissociable. In the standard PC framework, an increase in precision (i.e., a more concentrated, less uncertain policy) is typically accompanied by a change in the relative weighting of reward maximization, 
$Q(s,a)$, and the marginal action distribution, $P(a)$. In the present extension, however, these two aspects are no longer tightly coupled. As a result, the agent may become more or less precise independently of whether it places greater weight on reward or the marginal distribution.

\section{The Challenge of Computing the Optimal Policy}\label{challenge_section}

The attentive reader may have noticed that the optimal-policy equations in both the original PC framework (Eq. \eqref{eq:optimal_policy_compression}) and the present extension (Eq. \eqref{eq:policy_extended}) are implicitly circular. In both cases, the expression $\pi^{*}(a \mid s)$ depends on the marginal action distribution $P^{*}(a)$, while $P^{*}(a)$ is itself defined in terms of $\pi^{*}(a \mid s)$. How, then, can the optimal policy be computed?

In the PC framework, the optimal solution can be computed using a relatively simple iterative procedure known as the Blahut-Arimoto (BA) algorithm \cite{arimoto1972algorithm,blahut1972computation}. Starting from an initial guess for the marginal action distribution $P(a)$, one alternates between two steps: first, updating the conditional policy $\pi(a\mid s)$ using the current value of $P(a)$; and second, recomputing the marginal distribution as $P(a)=\sum_s p(s)\pi(a\mid s)$ using the updated policy. Repeating these two steps yields a sequence of policies that is guaranteed to converge to the optimal solution. This global-optimum guarantee holds because the objective consists of a linear reward term minus a mutual-information penalty, resulting in a concave optimization problem.

Thus, fitting a model based on the PC framework to behavioural data requires solving two coupled optimization problems. For any candidate pair of parameters ($\beta$, $\eta$), one must first solve an inner optimization problem to obtain the corresponding policy and, from it, the likelihood of the observed choices. Once this likelihood has been computed, an outer optimization can then be performed to identify the parameter values that maximize it. In other words, fitting the model amounts to solving a \textit{bilevel} (or \textit{nested}) optimization problem. 

In the present work, we have not conducted the analysis required to determine whether a BA-style algorithm could be used to monotonically increase the objective and converge to a local optimum. However, even if such an iterative procedure could be applied, it would no longer inherit the global-optimum guarantee of the original framework. The reason is that adding a penalty on conditional entropy breaks the concavity of the objective. As a result, even if an algorithm analogous to BA were developed to solve the present optimization problem, it would likely require additional measures--such as multiple random initializations--to reduce sensitivity to poor local solutions in a potentially complex optimization landscape.

\section{Conclusion and Future Work}

In this work, we argue that the definition of cognitive cost in the standard PC framework may be incomplete because it disregards the potential cost of irreducible uncertainty--that is, the policy uncertainty given that we know the state. The assumption that such uncertainty is costless is difficult to reconcile with evidence that reaction times are influenced not only by the extent to which a policy deviates from the marginal action distribution, but also by action uncertainty within a state. To address this, we propose an extension of the PC framework in which cognitive cost depends on both policy complexity and conditional entropy.

The proposed extension has two main implications. First, it introduces a new parameter, $\eta$, that increases the flexibility with which policy precision can vary without altering the relative balance between reward maximization and adherence to the marginal action distribution, unlike in the original framework, in which increasing precision and changing that balance cannot be decoupled. This added flexibility may help capture human behaviour more accurately, although future empirical work must test whether the increase in model complexity is justified. Second, the extended model predicts that the standard PC framework may underestimate cognitive costs, which could help explain the systematic observation that participants tend to deviate from the optimal reward-complexity trade-off (although other possible explanations, such as incomplete learning, have been proposed \cite{lai2021policy}).

At the same time, we acknowledge that the extended model introduces a practical challenge for fitting the model to human data, because, unlike in the standard PC framework, the optimal policy can no longer be computed straightforwardly due to the added complexity of the objective function.

Future work should therefore focus on developing an optimization procedure for fitting our proposed extended model. Such a procedure could be validated through a parameter recovery analysis \cite{wilson2019ten}, in which simulated or \textit{synthetic} behavioural datasets are generated using known parameter values, the model is then fitted to those synthetic datasets, and the new or \textit {recovered} parameter estimates are compared with the true generating values to determine whether the model and fitting procedure can produce reliable parameter estimates. Once this is achieved, our extended model could be compared with the standard PC framework using existing public datasets on contextual multi-armed bandit tasks (e.g., \cite{mcdougle2021modeling, lai2024human}).

If supported empirically, our work may have implications not only for the PC framework, but also for other frameworks that model decision-making as a trade-off between reward maximization and information-processing costs, such as active inference \cite{smith2022step} and capacity-limited reinforcement learning \cite{arumugam2024bayesian}. Indeed, all of these frameworks derive an optimal policy equation that takes the same form as the optimal solution in the PC framework (Eq. \eqref{eq:optimal_policy_compression}) \cite{sweeney2025decision}.

More broadly, our extension may also be relevant to research that investigates the benefits of modelling human decision-making as boundedly rational in order to inform the design of AI systems that interact effectively with humans \cite{callaway2022leveraging}. A deeper understanding of the mechanisms that shape cognitive cost may not only support the design of such systems, but also contribute to a better understanding of emerging behavioural phenomena such as cognitive offloading to large language models and related tools \cite{armitage2025nature, risko2016cognitive}.

\section*{Appendix}

Let $s \in S$ be the set of states and $a \in A$ the set of possible actions. 

We consider the following optimization problem:
\begin{equation}
\max_{\pi}\;\; V^{\pi}
\label{apx:eq:main_obj}
\end{equation}
subject to
\[
I(S;A)+ \eta \, H(A\mid S)= C
\]

Where $V^{\pi}$ is the expected value of a given policy:
\[
V^{\pi}
=\sum_s P(s)\sum_a \pi(a\mid s)\,Q(s,a)
\]

The mutual information (or policy complexity) is:
\[
I(S;A)
=\sum_s P(s)\sum_a \pi(a\mid s)\log\frac{\pi(a\mid s)}{P(a)}
\]

Where $P(a)=\sum_s P(s)\pi(a|s)$ is the marginal probability of choosing action $a$.

Finally, the conditional entropy is given by:
\[
H(A\mid S)=-\sum_s P(s)\sum_a \pi(a\mid s) \log \pi(a \mid s)
\]

Next, we expand the policy complexity term:

\begin{align*}
I(S;A)&=\sum_s P(s)\sum_a \pi(a\mid s)\log \pi(a\mid s)-\sum_{s}P(s)\sum_a \pi(a\mid s)\log P(a)\\
&=\sum_s P(s)\sum_a \pi(a\mid s)\log \pi(a\mid s)-\sum_{a}\sum_s P(s)\pi(a\mid s)\log P(a)\\
&=\sum_s P(s)\sum_a \pi(a\mid s)\log \pi(a\mid s)-\sum_{a} P(a) \log P(a)
\end{align*}

The objective in \eqref{apx:eq:main_obj} can be solved by first deriving the corresponding Lagrangian and introducing the multipliers $\beta$ (for reward maximization) and $\lambda_s$ (for normalization):

\begin{equation}
\mathcal L(\pi)
=
\beta \, V^{\pi}-[I(S;A)+ \eta \,H(A\mid S)]
+\sum_s \lambda_s\left(\sum_a \pi(a\mid s)-1\right)
\label{apx:eq:lagrangian}
\end{equation}

We now expand the form of the Lagrangian:

\begin{align*}
\mathcal{L}(\pi)
&=\beta\sum_s P(s)\sum_a \pi(a\mid s)Q(s,a)-
\sum_s P(s)\sum_a \pi(a\mid s)\log \pi(a\mid s)
+\sum_a P(a)\log P(a)
 \\
&-\eta\left[-\sum_s P(s)\sum_a \pi(a\mid s) \log \pi(a \mid s)\right] +\sum_s \lambda_s\left(\sum_a \pi(a\mid s)-1\right)\\
&=\beta\sum_s P(s)\sum_a \pi(a\mid s)Q(s,a)
+(\eta-1)\sum_s P(s)\sum_a \pi(a\mid s)\log \pi(a\mid s)\\
&+\sum_a P(a)\log P(a) +\sum_s \lambda_s\left(\sum_a \pi(a\mid s)-1\right)
\end{align*}

To find the solution we need to differentiate Eq. \eqref{apx:eq:lagrangian} w.r.t. $\pi(a\mid s)$ and set it equal to zero:

\begin{equation}
    \frac{\partial \mathcal L(\pi,\lambda)
}{\partial \pi(a\mid s)} = 0
\end{equation}

We compute the derivative term by term.

The reward term:
\[
\frac{\partial}{\partial \pi(a\mid s)}
\left[\beta \sum_{s^{\prime}} P(s^{\prime})\sum_{a^{\prime}} \pi(a^{\prime}\mid s^{\prime})Q(s^{\prime},a^{\prime})\right]
=
\beta\,P(s)\,Q(s,a)
\]

Next, we differentiate the second term of $\mathcal{L}(\pi)$ using the property $\frac{d}{dx}(x\log x)=\log x+1$:
\[
\frac{\partial}{\partial \pi(a\mid s)}
\left[(\eta-1)\sum_{s^{\prime}}P(s^{\prime})\pi(a^{\prime}\mid s^{\prime})\log \pi(a^{\prime}\mid s^{\prime})\right]
=
(\eta -1)P(s)\big(\log \pi(a\mid s)+1\big)
\]

Next, we differentiate the third term of the Lagrangian. We recall:
\[
P(a)=\sum_{s'} P(s')\pi(a\mid s'),
\qquad
\frac{\partial P(a)}{\partial \pi(a\mid s)}=P(s)
\]
Hence, using the chain rule:
\[
\frac{\partial}{\partial \pi(a\mid s)}\big[P(a)\log P(a)\big]
=\frac{\partial P(a)}{\partial \pi(a\mid s)}\frac{\partial}{\partial P(a)}\big[P(a)\log P(a)\big]
= P(s)\big(\log P(a)+1
\big)\]
Therefore,
\[
\frac{\partial}{\partial \pi(a\mid s)}
\left[\sum_{a^{\prime}} P(a^{\prime})\log P(a^{\prime})\right]
=
P(s)\big(\log P(a)+1\big)
\]

Finally, we differentiate the constraint term:
\[
\frac{\partial}{\partial \pi(a\mid s)}
\left[\sum_{s'}\lambda_{s'}\left(\sum_{a'}\pi(a'\mid s')-1\right)\right]
=\lambda_s
\]

Putting everything together:
\[
\frac{\partial \mathcal L}{\partial \pi(a\mid s)}
=
\beta P(s)Q(s,a)
+P(s)(\eta -1)\big(\log \pi(a\mid s)+1\big)
+P(s)\big(\log P(a)+1\big)
+\lambda_s
\]

Setting it to zero and dividing by $P(s)$ (assuming $P(s)>0$ for relevant states):

\[
0=
\beta \,Q(s,a)+(\eta-1)(\log \pi(a\mid s)+1)+\log P(a)+1+\frac{\lambda_s}{P(s)}
\]

Rearrange:
\[
(1- \eta)\log \pi(a\mid s)
=
\beta \, Q(s,a)+\log P(a)
+\underbrace{\left(\eta+\frac{\lambda_s}{P(s)}\right)}_{\text{constant in }a}
\]

For fixed state $s$, the bracketed term does not depend on $a$. Denote it by $k_s$. Then
\[
\log \pi(a\mid s)=\frac{1}{1-\eta} \left[ \beta \, Q(s,a)+\log P(a)+k_s\right] 
\]

Exponentiating:
\[
\pi(a\mid s)
=
\exp \left(c_s\right)
\exp \left\{\frac{1}{1-\eta}[\beta Q(s,a)+\log P(a)]\right\} \]

Let
\[
Z(s)=\sum_{a'} \exp \,\left\{ \frac{1}{1- \eta}[\beta \,Q(s,a')+\log P(a')]\right\}
\]
and absorb $\exp{(k_s)}$ into normalization, giving the optimal policy form:
\begin{equation}
\boxed{
\pi^*(a\mid s)=
\frac{
\exp\,\left\{\frac{1}{1-\eta}[\beta \,Q(s,a)+\log P^*(a)]\right\}}{Z(s)}}
\label{apx:eq:policy_exp_form}
\end{equation}
with
\begin{equation}
\boxed{
P^*(a)=\sum_s p(s)\pi^*(a\mid s)
}
\label{apx:eq:marginal_fixed_point}
\end{equation}

\bibliographystyle{vancouver} 
\bibliography{references}

\end{document}